\documentclass[aps,pra,twocolumn,floatfix,showpacs,superscriptaddress]{revtex4}
\usepackage[dvipdfm]{graphicx}
\usepackage{amssymb}
\usepackage{amsmath}

\begin{document}
\title{Chaos-assisted emission from asymmetric resonant cavity microlasers}

\newcommand{\MPIPKS}{\affiliation{Max-Planck-Institut f\"ur Physik
  Komplexer Systeme, N\"othnitzer Stra\ss e 38, D-01187 Dresden,
  Germany}}
\newcommand{\OPU}{\affiliation{Department of Communication
    Engineering, Okayama Prefectural University, 111 Kuboki, Soja,
    Okayama 719-1197, Japan}}
\newcommand{\NTTCS}{\affiliation{NTT Communication Science
Laboratories, NTT Corporation, 2-4 Hikaridai, Seika-cho, Soraku-gun,
Kyoto 619-0237, Japan}}
\newcommand{\PURDUE}{\affiliation{Birck Nanotechnology Center,
    Department of Electrical and Computer Engineering, Purdue
    University, 1205 West State Street West Lafayette, Indiana
    47907-2057, USA}}

\author{Susumu Shinohara}\MPIPKS
\author{Takahisa Harayama}\NTTCS
\author{Takehiro Fukushima}\OPU
\author{Martina Hentschel}\MPIPKS
\author{Satoshi Sunada}\NTTCS
\author{Evgenii E. Narimanov}\PURDUE

\newcommand{\degrees}{^{\circ}}
\newcommand{\micron}{\mu\mbox{m}}
\newcommand{\etal}{{\it et al.}}

\begin{abstract}
We study emission from quasi-one-dimensional modes of an asymmetric
resonant cavity that are associated with a stable periodic ray orbit
confined inside the cavity by total internal reflection.
It is numerically demonstrated that such modes exhibit directional
emission, which is explained by {\it chaos-assisted emission} induced by
dynamical tunneling.
Fabricating semiconductor microlasers with the asymmetric resonant
cavity, we experimentally demonstrate the selective excitation of the
quasi-one-dimensional modes by employing the device structure to
preferentially inject currents to these modes and observe directional
emission in good accordance with the theoretical prediction based on
chaos-assisted emission.
\end{abstract}


\maketitle

\section{Introduction}

Besides growing interest from the fundamental physics of light and
practical applications in photonics
\cite{Yamamoto93,Chang96,Vahala03,Vahala04}, two-dimensional optical
microcavities have attracted much attention from the field of
quantum/wave chaos, since they can be regarded as physical realization
of dynamical billiards, a well-studied model in the field (for reviews
see for example Refs. \cite{Nockel96,Schwefel04,Harayama11}).
Concepts of the quantum/wave chaos theory have been successfully
applied to two-dimensional microcavities to explain how ray chaos
influences the characteristics of resonant modes such as emission
directionality and lifetimes.

One of such applied concepts is {\em dynamical tunneling}
\cite{Davis81}, which refers to tunneling between classical invariant
components disconnected from each other under classical dynamics (see
Ref. \cite{Creagh98} for a review).
In generic Hamiltonian systems, i.e., those with mixed phase space,
where regular and chaotic dynamics coexist, dynamical tunneling
manifests itself in the fundamental processes such as
regular-to-chaotic and chaotic-to-regular tunneling
\cite{Hanson84,Shudo95,Sheinman06,Baecker08}.
One of the dramatic effects of dynamical tunneling is observed in the
enhancement of level splitting of doublets associated with
symmetry-related regular components separated by a global chaotic sea.
The phenomenon is explained by the mechanism known as {\em
chaos-assisted tunneling}
\cite{Bohigas93,Tomsovic94,Doron95,Leyvraz96,Frischat98,Lin90}; the
regular components are coupled not only by direct regular-to-regular
tunneling but also by a multi-step process involving
regular-to-chaotic and chaotic-to-regular tunneling, and this process
leads to a significant enhancement of the coupling.
The dynamical tunneling and chaos-assisted tunneling have been
experimentally observed in microwave billiards
\cite{Dembowski00,Baecker08b} and cold atom systems
\cite{Hensinger01,Steck01}.

In the context of optical microcavities, the effect of dynamical
tunneling was already pointed out in the pioneering theoretical study of
asymmetric resonant cavities \cite{Nockel97}, where the significant
shortening of resonance lifetimes is attributed to dynamical tunneling.
This effect was quantitatively studied by effective Hamiltonian models
\cite{Hackenbroich97,Podolskiy05,Narimanov06}, where a formula is
derived describing how resonance lifetimes depend on the factors such
as the area of regular components and wavenumber.
Strong influence of dynamical tunneling on the emission pattern of a
ray-chaotic cavity was also pointed out in
Refs. \cite{Podolskiy05,Narimanov06}.
Resonance lifetimes were connected with a direct regular-to-chaotic
tunneling rate in Ref. \cite{Baecker09}.
The enhancement and fluctuation of frequency splitting of regular mode
doublets is explained by the chaos-assisted tunneling in
Ref. \cite{Tureci02}.

Recently, the manifestation of dynamical tunneling was experimentally
observed in the emission patterns of semiconductor microlasers
\cite{Shinohara10,Stone10} and the pump-coupling efficiency of dye-doped
liquid jet microlasers \cite{Yang10}.
In the former study, an asymmetric resonant cavity was designed to have
quasi-one-dimensional modes strongly localized along a stable periodic
ray orbit that is confined by total internal reflection (TIR).
In the experiment, such modes were excited by selective pumping, and
highly directional emission was observed.
The purpose of this paper is to present a comprehensive study on this
microlaser.
The emission directionality of this microlaser was interpreted by the
following two steps: (i) dynamical tunneling from the TIR-confined ray
orbit to a neighboring chaotic ray orbit, and (ii) ray-dynamical
propagation following the chaotic dynamics leading to refractive
emission.
In this paper, we shall call this process {\em chaos-assisted
emission} in order to emphasize that the process is
regular-to-chaotic-to-continuum.

This paper is organized as follows: In Sec. \ref{sect:cavity}, we
introduce the asymmetric resonant cavity and explain its ray dynamics,
where particular attention is paid to the rectangle-shaped stable ray
orbit confined by TIR.
Section \ref{sect:rect-modes} presents the numerical study of the
resonant modes associated with this orbit (rectangle-orbit modes, in
short), showing that these quasi-one-dimensional modes exhibit
chaos-assisted directional emission.
Section \ref{sect:device} explains the device structure of fabricated
semiconductor microlasers, especially focusing on the structure to
achieve selective excitation of the rectangle-orbit modes.
Section \ref{sect:spectra} provides measured spectral data that evidence
the selective excitation.
Section \ref{sect:emission} presents measured emission patterns which
are attributed to the chaos-assisted emission.
Finally, our conclusion is given in Sect. \ref{sect:conclusion}.

\section{Asymmetric resonant cavity}
\label{sect:cavity}
\begin{figure}[t]
\includegraphics[width=0.48\textwidth]{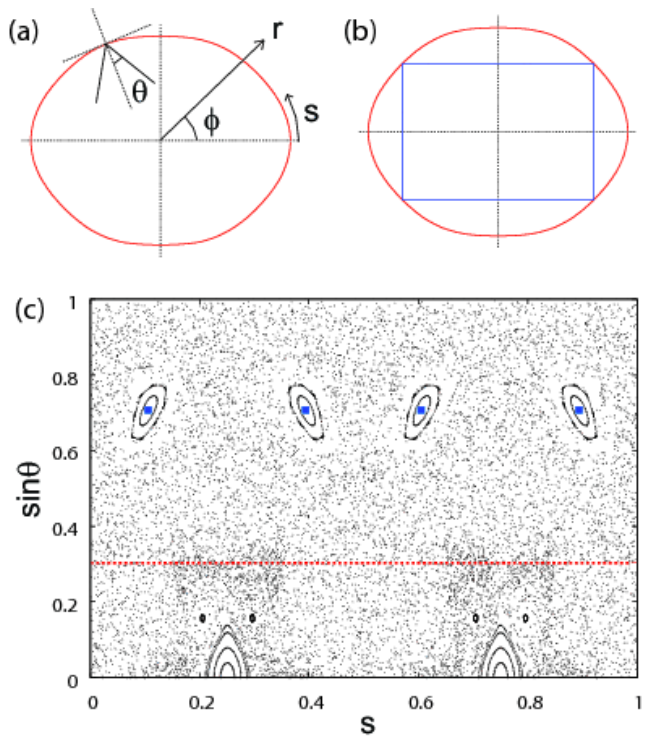}
\caption{(color online) (a) The geometry of the asymmetric resonant
cavity with the definitions of the polar coordinates $(r,\phi)$ and of
the Birkhoff coordinates $(s,\sin\theta)$. (b) The rectangular ray
orbit. (c) The upper half of the Poincar\'e surface of section (PSOS)
for the ray dynamics. The coordinate $s$ is normalized by the perimeter
of the cavity. The critical line for total internal reflection,
$\sin\theta=1/3.3$, is plotted by a dotted line. The four filled squares
($\blacksquare$) represent the rectangular orbit shown in (b).}
\label{fig:cavity}
\end{figure}

In this paper, we consider the asymmetric resonant cavity whose shape is
defined in the polar coordinates $(r,\phi)$ as
\begin{equation}
r=R(\phi)=r_0 (1+a\cos 2\phi+b\cos 4\phi+c\cos 6\phi),
\label{eq:cavity}
\end{equation}
where $r_0$ is the size parameter and $a, b$, and $c$ are deformation
parameters fixed as $a=0.1, b=0.01$, and $c=0.012$.
The geometry of the cavity is depicted in Fig. \ref{fig:cavity} (a).

An important feature of this cavity is the existence of the 4-bounce
stable ray orbit drawing the ``rectangle'' as shown in
Fig. \ref{fig:cavity} (b), which is confined by TIR when the
refractive index $n>\sqrt{2}$.
The peculiarity of the rectangular orbit in relation with global ray
dynamics is revealed by plotting the Poincar\'e surface of section
(PSOS) of the ray dynamics.
The PSOS describes the successive bounces of a ray trajectory usually
using the Birkhoff coordinates $(s,\sin\theta)$, where $s$ represents
the arclength along the cavity and $\theta$ the incident angle as
illustrated in Fig. \ref{fig:cavity} (a).
The PSOS is convenient for describing the openness of the dielectric
cavity, because the condition for TIR is given by $|\sin\theta|>1/n$.
Throughout this paper, we assume $n=3.3$.
This value corresponds to the effective refractive index calculated
for the structure of actually fabricated devices to be explained in
Sect. \ref{sect:device}.

Because the PSOS is symmetric under $\sin\theta\mapsto -\sin\theta$, we
show only the upper half of the PSOS in Fig. \ref{fig:cavity} (c), where
one finds that ray dynamics is mostly chaotic, and the four islands of
stability corresponding to the rectangular orbit are the only dominant
islands located in the TIR-region, i.e., $\sin\theta>1/3.3$.
In this paper, we focus on the resonant modes that are strongly
localized on the TIR-confined dominant islands.
The ray-dynamical property that these islands are isolated and only
dominant islands in the global chaotic sea will later enable us to
interpret the emission patterns of those resonant modes by a simple
scenario based on chaos-assisted emission.

\section{Rectangle-orbit modes}
\label{sect:rect-modes}
Resonant modes of the cavity are given as the solutions of the
Helmholtz equation derived from Maxwell's equations
\cite{Nockel97,Nockel96,Schwefel04}.
They are quasibound states with finite lifetimes characterized by
complex wavenumbers $k$.
In this paper, we are interested in quasi-one-dimensional modes
strongly localized along the stable rectangular orbit.
The existence of such modes can be proven in the short-wavelength
limit by the so-called Gaussian optics \cite{Tureci02}.
This type of modes has been previously studied, for instance, for the
so-called 'bow-tie' laser \cite{Gmachl98}, but an important difference
here is that the corresponding ray orbit is confined by TIR.
Thus, the ray optics based on Fresnel's law as well as the Gaussian
optics predict that rectangle-orbit modes are perfectly confined,
which is not true for accurate wave description.
In order to correctly capture the emission patterns of the
rectangle-orbit modes, below we resort to numerics.

\begin{figure}[b]
\includegraphics[width=0.48\textwidth]{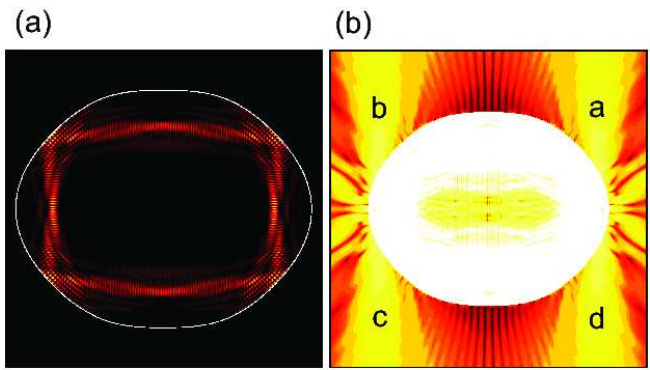}
\caption{(color online) (a) Field intensity distribution of the
  rectangle-orbit mode with the wavenumber $k=49.94-i\,0.00012$. (b)
  Log-scale plot of the rectangle-orbit mode. The intensity increases
  as the color changes from black to white.}
\label{fig:wfunc}
\end{figure}

An accurate way for numerically calculating the modes is provided, for
instance, by the boundary element method \cite{Wiersig03}.
In Fig. \ref{fig:wfunc} (a) we present a result obtained by this
method, showing the field intensity distribution of a resonant mode
strongly localized along the rectangle orbit, where one can see a
regular modal pattern along the orbit.
Its wavenumber is $k=49.94-i\,0.00012$.
Concerning the resonant mode calculation, we note that the average
radius of the cavity is fixed as $r_0=1$.
We employ the boundary condition for the transverse-electric
polarization, to be consistent with experimental data presented later.

In order to observe exponentially small emission from the
rectangle-orbit mode, we plot the field intensity distribution in log
scale in Fig. \ref{fig:wfunc} (b), where one finds that dominant
emission occurs on both sides of the cavity towards the directions
$\phi=\pm 90\degrees$.
\begin{figure}[b]
\includegraphics[width=0.48\textwidth]{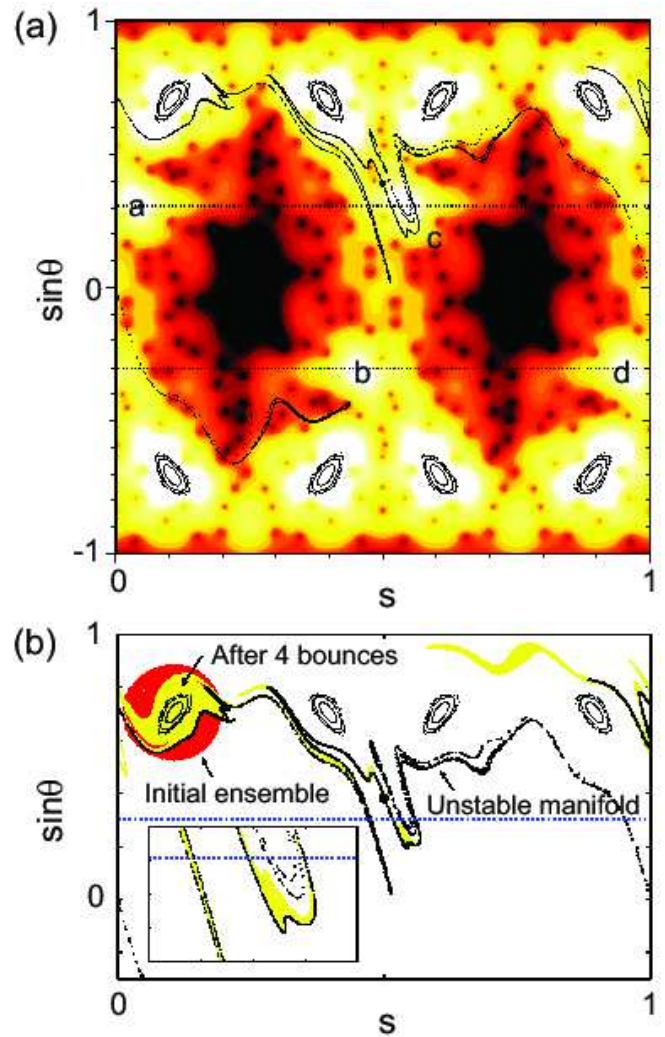}
\caption{(color online) (a) Log-scale plot of the Husimi distribution
for the rectangle-orbit mode with $k=49.94-i\,0.00012$, where the
islands corresponding to the rectangular orbit and the unstable
manifolds of the unstable period-3 point $(0.5, 0.389...)$ are
superimposed. The intensity increases as the color changes from black to
white. The critical lines for total internal reflection, $\sin\theta=\pm
1/3.3$, are plotted by dotted lines. (b) Ray-dynamical evolution of an
ensemble initially prepared over the leftmost island and its
neighborhood in the upper half PSOS. The inset shows a magnification
around the area where the unstable manifold overlaps with the critical
line.}  \label{fig:husimi}
\end{figure}
These emissions are understood as the wave tunneling effect intrinsic
to a ray-chaotic cavity.
This emission mechanism is better explained by the PSOS than the real
space.
Thus, we consider the PSOS-representation of an eigenfunction called
the Husimi distribution \cite{Crespi93,Hentschel03}.
Figure \ref{fig:husimi} (a) shows the Husimi distribution of the
rectangle-orbit mode shown in Fig. \ref{fig:wfunc} (a), where it is
plotted in log scale to see exponentially small intensities in the
area of the chaotic sea.
In Fig. \ref{fig:husimi} (a), the intensity increases as the color
changes from black to white.
Reflecting the fact that the mode is localized along the rectangular
orbit, we find high intensities on the corresponding islands.
In addition, we find some intensities spread over the chaotic sea,
even reaching down to the critical lines for TIR,
$\sin\theta=\pm 1/3.3$.
The four high-intensity spots on the critical line labeled $a, b, c$,
and $d$ correspond to the four dominant emissions observed in
Fig. \ref{fig:wfunc} (b).
Together with the field intensity distribution in the real space
(Fig. \ref{fig:wfunc} (b)), the Husimi distribution reveals that
emission occurs at two close but different boundary points (one
towards $\phi=90\degrees$ and the other towards $\phi=-90\degrees$) on
each side of the cavity.

The emission pattern of the rectangle-orbit mode can be qualitatively
explained in the following manner \cite{Podolskiy05,Narimanov06}.
Although intensities are highly localized on the islands corresponding
to the rectangular orbit, some intensities are leaked to their
neighboring chaotic orbits via dynamical tunneling.
Once this happens, the leaked intensities are transported to the leaky
region (i.e., $|\sin\theta|<1/3.3$) by the chaotic dynamics, resulting
in refractive emission.
Hence, it is the ray dynamics around the critical lines for TIR that
determines the emission pattern.
In fact, such ray dynamics is governed by the unstable manifolds
emanating from a short-periodic unstable periodic point located near
the critical line
\cite{Schwefel04b,SY.Lee05,Shinohara06,SB.Lee07,Wiersig08}.
In Fig. \ref{fig:husimi} (b), we show how a ray ensemble initially
prepared around the leftmost island is mapped to after 4 bounces.
The initially circular area is stretched and folded after 4 bounces,
partially penetrating into the region below the critical line as
magnified in the inset of Fig. \ref{fig:husimi} (b).
As illustrated in Fig. \ref{fig:husimi} (a), the high-intensity spot
labeled $c$ coincides with the position where the unstable manifold
crosses the critical line.
The positions of the other high-intensity spots ($a, b$ and $d$) are
explained in the same manner.

In the remaining part of this paper, we study whether the
theoretically predicted chaos-assisted emission from the
rectangle-orbit modes can be experimentally observed.
Therefore, it is crucial in experiments to excite only the
rectangle-orbit modes, suppressing the excitation of the other modes.
In order to achieve this selective excitation, we employ the selective
pumping method that has been previously examined for
quasi-stadium-shaped microcavity lasers, where the selective excitation
of two different types of modes, called the axis modes and ring modes,
has been successfully demonstrated
\cite{Harayama03,Fukushima04a,Fukushima04b,Fukushima05,Choi06,Fukushima07,Choi08}.

\section{Device structure}
\label{sect:device}
\begin{figure}[b]
\includegraphics[width=0.48\textwidth]{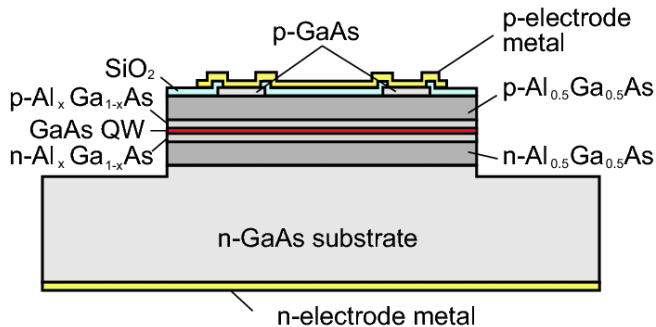}
\caption{(color online) Schematic cross-sectional view of the device.}
\label{fig:structure}
\end{figure}
\begin{figure}[b]
\includegraphics[width=0.48\textwidth]{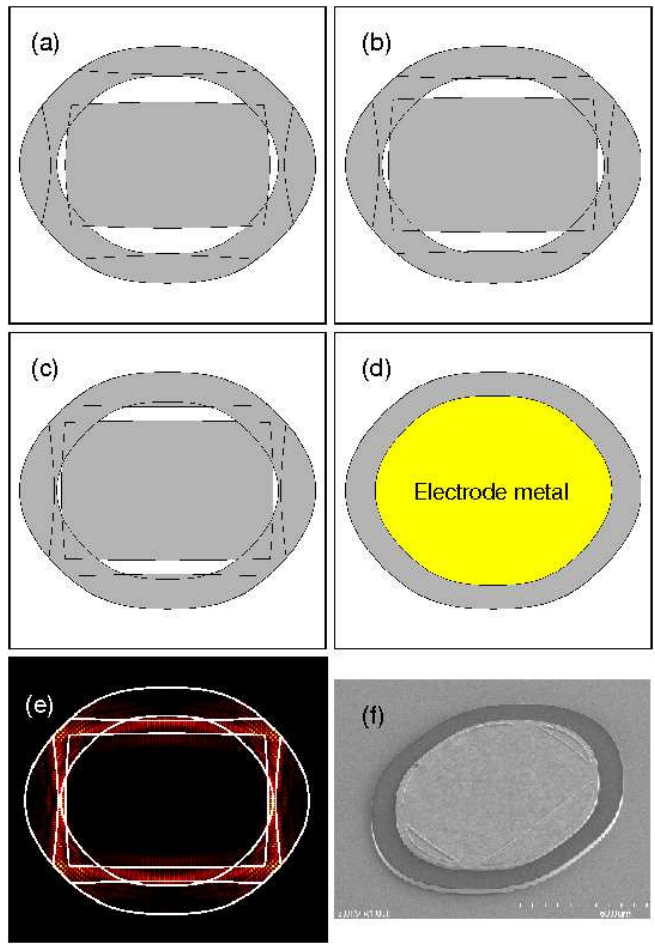}
\caption{(color online) (a)-(c): Top-view geometry of the SiO$_2$
  layer. The white regions inside the cavity define the contact window
  where SiO$_2$ is removed and through which currents are injected.
  The contact window is broad in (a), intermediate in (b), and narrow
  in (c). We make the ``margin area'' on the rim, whose shape is
  defined by $0.75 R(\phi)\leq r \leq R(\phi)$. Above the SiO$_2$
  layer, the $p$-electrode metal is deposited as in (d), whose shape
  is defined by $r\leq 0.8 R(\phi)$. (e) Field intensity distribution
  of the rectangle-orbit mode with $k=49.94-i0.00012$ superimposed
  with the narrow contact window. (f) Scanning electron microscope
  image of a fabricated device with the broad contact window. The
  average radius is $r_0=50$ $\micron$.}
\label{fig:contact}
\end{figure}

In this section, we explain the device structure to achieve the
selective excitation (see Ref. \cite{Fukushima04b} for the details of
the fabrication process).
Figure \ref{fig:structure} shows a schematic cross-sectional view of
the device.
The device is fabricated from MOCVD-grown material with a
1.5-$\micron$ $n$-Al$_{0.5}$Ga$_{0.5}$As lower cladding layer, a
0.2-$\micron$ $n$-Al$_x$Ga$_{1-x}$As ($x=0.5-0.2$) graded region, a
10-nm GaAs quantum well layer, a 0.2-$\micron$
$p$-Al$_x$Ga$_{1-x}$As($x=0.2-0.5$) graded region, a 1.5-$\micron$
$p$-Al$_{0.5}$Ga$_{0.5}$As upper cladding layer, a 0.2-$\micron$
$p$-GaAs contact layer, and a 400-nm SiO$_2$ layer.
The average radius $r_0$ of the device is fixed as 50 $\micron$.

The key structure for the selective excitation is the ``contact window''
etched to the SiO$_2$ layer.
The SiO$_2$ layer works as the insulation between the semiconductor
layers and the $p$-electrode metal, enabling a selective current
injection through the contact window.
The top-view shape of the $p$-GaAs layer is given by adding
1-$\micron$ margins to the regions of the contact window which is
formed by a dry etching process before depositing the SiO$_2$ layer.

In order to excite only the rectangle-orbit modes, we form the contact
window along the rectangular orbit as illustrated in
Figs. \ref{fig:contact} (a)-(c).
It is not known a priori what design of the contact window is most
optimal for stable and efficient excitation of the rectangle-orbit
modes.
To test the sensitivity of the selective excitation on a contact window
pattern, we examine three different patterns of the contact window,
i.e., a broad, an intermediate, and a narrow one, respectively shown in
Figs. \ref{fig:contact} (a), \ref{fig:contact} (b), and
\ref{fig:contact} (c).
We show in Fig. \ref{fig:contact} (e) how the narrow contact window
overlaps with the rectangle-orbit mode.

On the rim of the cavity, we make the ``margin area'' where SiO$_2$ is
left.
This is to avoid excitation of whispering-gallery-type modes that are
localized along the rim of the cavity.
The precise shape of the margin area is given by $0.75 R(\phi)\leq
r\leq R(\phi)$.
The $p$-electrode metal is deposited over the contact window and part
of the surrounding SiO$_2$ layer as shown in Fig. \ref{fig:contact}
(d) and its shape is defined by $r\leq 0.8 R(\phi)$.
In Fig. \ref{fig:contact} (f), we show the scanning electron
microscope image of a fabricated device with the broad contact window.

\section{Spectral characteristics}
\label{sect:spectra}
\begin{figure}[b]
\includegraphics[width=0.45\textwidth]{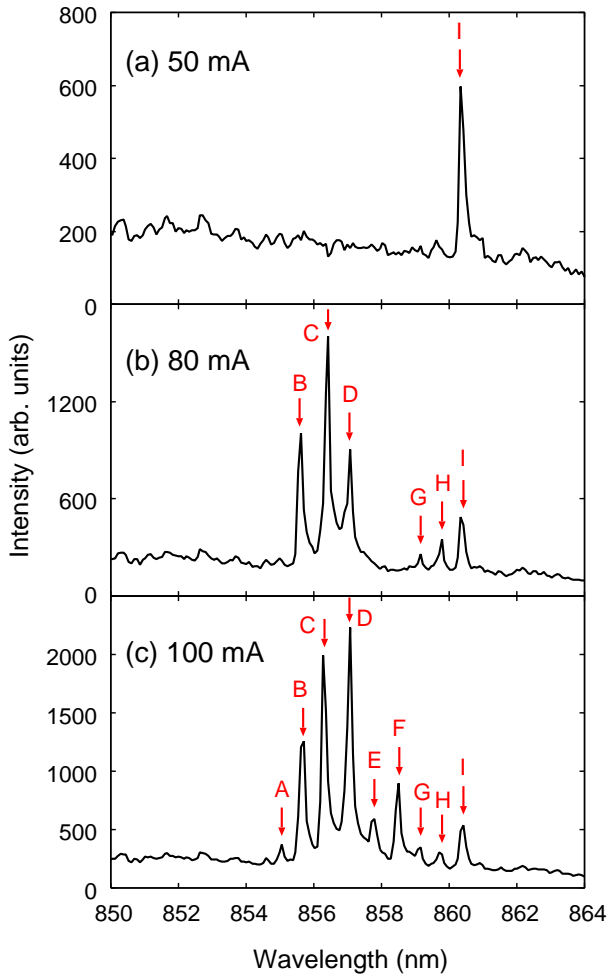}
\caption{(color online) Lasing spectra of the device with the narrow
  contact window for different pumping currents: (a) 50 mA; (b) 80 mA;
  (c) 100 mA. Equidistantly spaced peaks are observed in (c), where
  the average mode spacing is 0.67 nm, in good agreement with the
  longitudinal mode spacing of the rectangle-orbit modes, 0.63 nm.}
\label{fig:spectra.no185}
\end{figure}
As, for instance, the Gaussian optic theory directly demonstrates
\cite{Tureci02}, rectangle-orbit modes are regarded as the
quantization along the rectangular orbit.
Their longitudinal mode spacing is given by
\begin{equation}
\triangle\lambda=
\frac{\lambda^2}{n L \left(1-{\displaystyle \frac{\lambda}{n}\frac{dn}{d\lambda}}\right)},
\label{eq:mode-spacing}
\end{equation}
where $\lambda$ is the wavelength, $n$ the effective refractive index,
$L$ the path length of the rectangle orbit, and $dn/d\lambda$ the
dispersion.
In this section, we present measured spectral data showing
equidistantly spaced peaks corresponding to the above mode spacing,
which constitutes definitive evidence that rectangle-orbit modes only
are excited.
The devices are tested in pulsed mode operation at $25\degrees$C with
pulse width 500 ns and repetition rate 1 kHz.
In Fig. \ref{fig:spectra.no185}, we show lasing spectra of the device
with the narrow contact window for the pumping currents 50 mA, 80 mA,
and 100 mA.
For 50 mA, which is just above the lasing threshold current, one can
confirm single-mode lasing.
For 100 mA, the spectrum shows equidistantly spaced peaks with the
averaged mode spacing $\triangle\lambda=0.67$ nm.
Considering the uncertainty of the refractive index and the resolution
of the measurement, one can conclude that this value corresponds to
the longitudinal mode spacing of the rectangle-orbit modes estimated
as 0.63 nm.
This theoretical estimate is calculated from
Eq. (\ref{eq:mode-spacing}) with $\lambda=860$ nm, $n=3.3$,
$L=283.04~\mu$m, and $dn/d\lambda=-1.0~\mu\mbox{m}^{-1}$.
Here we use the value of the dispersion $dn/d\lambda$ from
Ref. \cite{Casey78}.

\begin{figure}[b]
\includegraphics[width=0.48\textwidth]{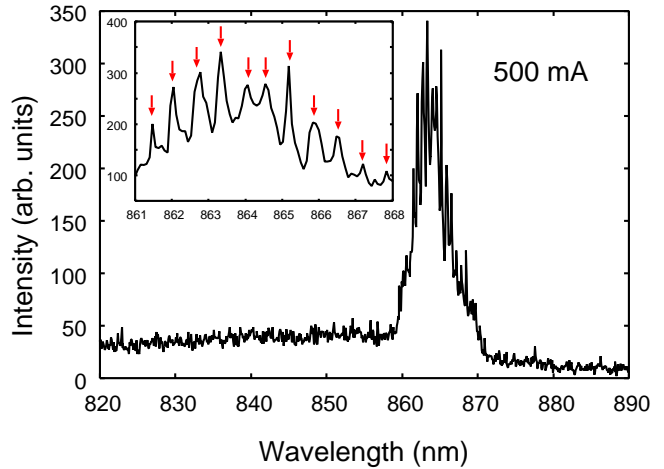}
\caption{(color online) Lasing spectrum of the device with the narrow
  contact window for the pumping current 500 mA. The inset shows the
  magnification of a high-intensity range, where equidistantly spaced
  peaks are observed, whose average mode spacing is 0.64 nm, in
  excellent agreement with the longitudinal mode spacing of the
  rectangle-orbit modes, 0.63 nm.}
\label{fig:spectra.no185.500mA}
\end{figure}

Because currents are not injected around the rim of the cavity, the
lasing of whispering-gallery-type modes is highly unlikely.
This can be confirmed by estimating the mode spacing corresponding to
whispering-gallery-type modes.
Using the above formula with $L=317.75~\mu\mbox{m}$ (i.e., perimeter
of the cavity), one can estimate their mode spacing as 0.56 nm, which
is considerably smaller than the measured mode spacing.

\begin{figure}[b]
\includegraphics[width=0.48\textwidth]{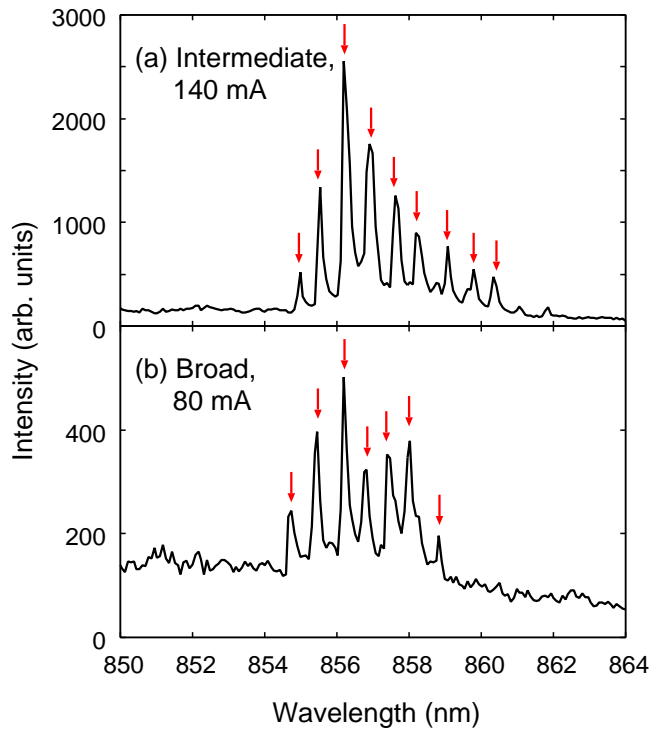}
\caption{(color online) Lasing spectra for (a) the device with the
  intermediate contact window at the pumping current 140 mA and for
  (b) the device with the broad contact window at 80 mA. Equidistantly
  spaced peaks are observed in both cases. The average mode spacing is
  0.67 nm in (a), while 0.68 nm in (b), both in good agreement with
  the longitudinal mode spacing of the rectangle-orbit modes, 0.63
  nm.}
\label{fig:spectra.182and179}
\end{figure}

The above theoretical estimates convince us that the equidistantly
spaced peaks in Fig. \ref{fig:spectra.no185} (c) are the
rectangle-orbit modes, not the whispering-gallery-type modes.
In addition, using Fig. \ref{fig:spectra.no185} (c) as a reference, we
can also conclude that peaks observed in Figs. \ref{fig:spectra.no185}
(a) and (b) correspond to the rectangle-orbit modes.

We confirmed selective excitation of the rectangle-orbit modes for a
much higher pumping current.
In Fig. \ref{fig:spectra.no185.500mA}, we show the lasing spectra for
the pumping current 500 mA, where one finds again equidistantly spaced
peaks with the average mode spacing 0.64 nm, which is in excellent
agreement with the theoretical estimate 0.63 nm.
The spectral peaks in Fig. \ref{fig:spectra.no185.500mA} are
red-shifted from those in Fig. \ref{fig:spectra.no185}, which is
considered to be due to a heat effect caused by the high pumping
current.

Concerning the lasing of higher transverse excited modes, we cannot find
its clear evidence for lower pumpings. However, for the pumping 500 mA,
we can identify subpeaks at, for instance, 861.7 nm and 862.4 nm, which
are considered to be higher transverse excitations.

We also confirmed that the selective excitation is achieved with the
intermediate and broad contact windows.
Figures \ref{fig:spectra.182and179} (a) and
\ref{fig:spectra.182and179} (b) show spectra for the device with the
intermediate contact window at 140 mA and for the device with the
broad contact window at 80 mA, respectively.
In both cases, one can confirm equidistantly spaced peaks.
The average mode spacing is 0.67 nm in
Fig. \ref{fig:spectra.182and179} (a) while 0.68 nm in
Fig. \ref{fig:spectra.182and179} (b).
These values are again in good agreement with the theoretical estimate
0.63 nm.
As for the lasing threshold currents, we could not find significant
dependence on the contact window patterns, i.e., the threshold currents
are around 50 mA irrespective of the contact window pattern.

\section{Emission characteristics}
\label{sect:emission}
For microcavities with mostly or fully chaotic ray dynamics, it has
been reported that experimental far-field patterns are well explained
by ray-dynamical calculations
\cite{Fukushima04b,SB.Lee07,Tanaka07,Shinohara08,Choi08b,Shinohara09}.
Generally, some intrinsic deviation is observed between the far-field
pattern of a {\em single} resonant mode and that of ray-dynamical
calculation, especially in the short-wavelength regime
\cite{Shinohara07,Shinohara08,Choi08b,Shinohara09}.
Nonetheless, it was demonstrated that the averaged far-field pattern
of many resonant modes is very well reproduced by ray-dynamical
calculation.
Because multi-mode lasing works as a mechanism to realize the
``average'' of many modes, one can expect that the correspondence
between experimental data and ray-dynamical calculation becomes the
better, the more the number of lasing modes increases.
In fact, the validity of this expectation was confirmed for a
stadium-shaped microlaser \cite{Choi08b}.
In this section, we report measured emission patterns of our devices
where the rectangle-orbit modes are selectively excited and show to
which extent they can be reproduced by ray-dynamical calculation.

\begin{figure}[t]
\includegraphics[width=0.48\textwidth]{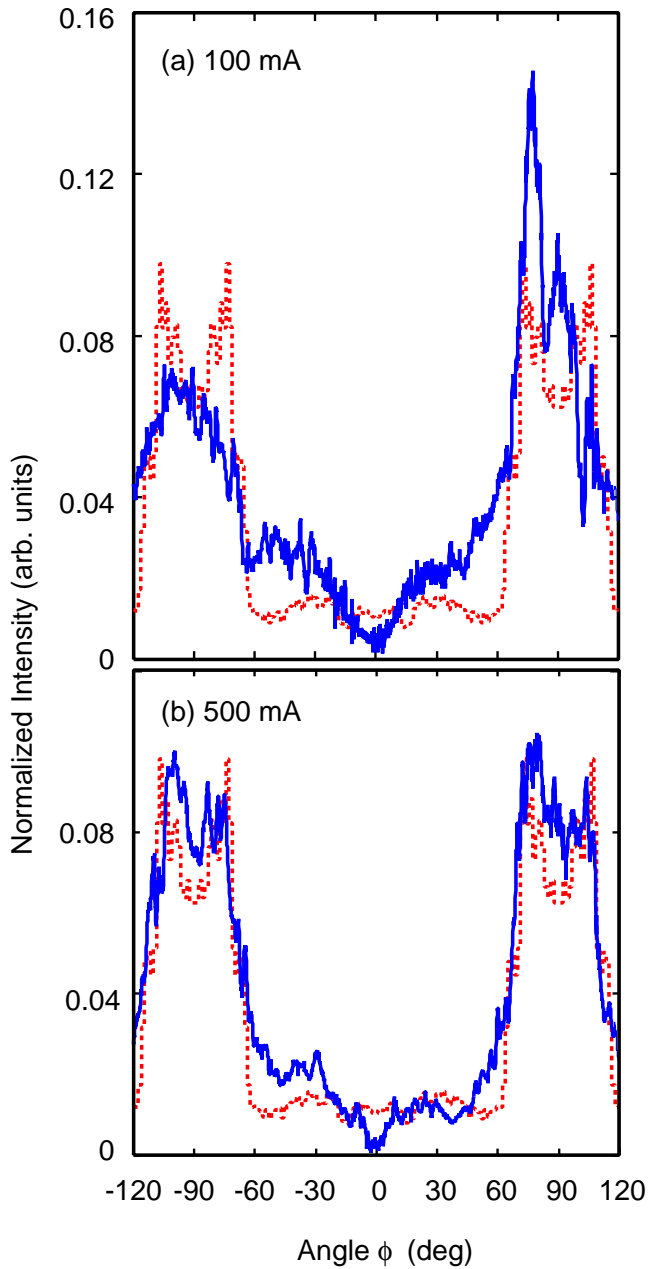}
\caption{(color online) Far-field emission patterns of the device with
the narrow contact window for (a) 100 mA and (b) 500 mA. The
definition of the angle $\phi$ is given in Fig. \ref{fig:cavity}
(a). The solid curves represent experimental data, while the dotted
curves represent ray-dynamical data.}
\label{fig:ffp}
\end{figure}

We show far-field patterns of the device with the narrow contact
window for the pumping currents 100 mA and 500 mA in
Figs. \ref{fig:ffp} (a) and \ref{fig:ffp} (b), respectively.
We note that these patterns are normalized after subtracting uniform
background due to spontaneous emission.
In good accordance with the theoretical prediction based on the
chaos-assisted emission, we observe strong emissions towards $\pm 90
\degrees$.

In Figs. \ref{fig:ffp} (a) and \ref{fig:ffp} (b), we superpose the
theoretical far-field pattern obtained by ray-dynamical calculation.
We find that the experimental far-field pattern for 500 mA is well
reproduced by ray-dynamical calculation even for the substructures of
the main peaks.
As shown in Ref. \cite{Shinohara10}, wave calculation of the far-field
pattern for a single mode does not show such substructures (see Fig. 5
of Ref. \cite{Shinohara10}).
Therefore we conclude that the substructures of the main peaks in the
measured far-field pattern are due to multimode lasing, because the
far-field pattern for ray-dynamical calculation can be considered as
that from an average of many modes as explained above.

For 100 mA, the experimental far-field pattern is asymmetric, showing
some deviations from the ray-dynamical calculation.
An intrinsic mechanism leading to the asymmetry is the locking of modes
belonging to different parities \cite{Harayama03,Sunada05}. 
The locking is caused by nonlinear modal interaction due to the lasing
medium, creating a lasing mode whose emission pattern violates the
symmetries of the cavity.
When the number of such a lasing mode is large, the asymmetry is
averaged out, resulting in a symmetric emission. 
However, the asymmetry becomes prominent when the number of such a mode
is relatively small \cite{Sunada05}.
For the pumping 500 mA, we can identify many modes with significant
intensities in the lasing spectrum shown in Fig. 7, while for the
pumping 100 mA we find only a couple of dominant lasing modes in the
lasing spectrum shown in Fig. 6 (c).
We expect that for 100 mA the relatively small number of the lasing
modes makes the emission asymmetry significant.

\begin{figure}[t]
\includegraphics[width=0.48\textwidth]{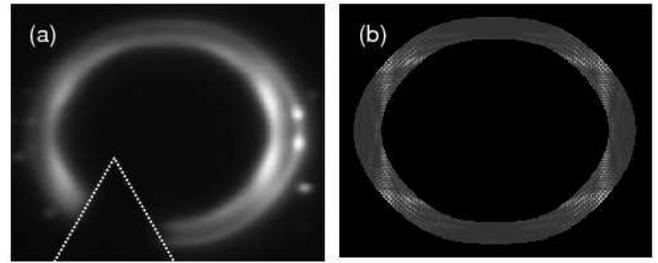}
\caption{(a) CCD photo of the lasing device with the broad contact
window for 300 mA. The photo is taken from a slightly oblique angle,
causing the difference between the right hand edge and left hand
edge. The outline of a needle for current injection is indicated by
the dotted line. (b) The field intensity distribution of the
rectangle-orbit mode shaded by the electrode metal area.}
\label{fig:nfp}
\end{figure}

Lastly, we show a CCD photo of the lasing device with the broad
contact window at 300 mA in Fig. \ref{fig:nfp} (a).
Although the lasing emission is in-plane, some light scattered inside
the cavity is observed from the top in the margin area.
For comparison, we show in Fig. \ref{fig:nfp} (b) the field intensity
distribution of the rectangle-orbit mode shaded by the electrode metal
area.
We can see that bright regions in Fig. \ref{fig:nfp} (a) are coming
mainly from the four corners of the rectangular orbit.
Moreover, we can observe two bright spots around the right hand edge
of the cavity.
These are considered to be the light scattered at the cavity boundary,
thus indicating the emission points.
These emission points are in good accordance with the theoretical
prediction drawn from the Husimi distribution shown in
Fig. \ref{fig:husimi} (a) as well as the near-field pattern shown in
Fig. \ref{fig:wfunc} (b).

\section{Conclusion}
\label{sect:conclusion}
We have focused on the quasi-one-dimensional modes of the asymmetric
resonant cavity, which are strongly localized along the stable
rectangular ray orbit that is confined by TIR.
Direct numerical calculation of such a resonant mode revealed that
directional emission occurs from cavity boundary points apart from the
rectangular ray orbit.
This at first sight counterintuitive result was explained by the
chaos-assisted emission mechanism consisting of regular-to-chaotic
tunneling and subsequent chaotic transport leading to refractive
emission.

We fabricated the cavity with the structure to selectively excite the
rectangle-orbit modes.
The success of the selective excitation was confirmed by the
observation of the equidistantly spaced peaks in the lasing spectra
well corresponding the longitudinal mode spacing of the
rectangle-orbit modes.
In the emission patterns, we observed directional emission attributed
to the chaos-assisted emission.

Our results demonstrated that the concepts of quantum/wave chaos and
nonlinear dynamics such as dynamical tunneling and unstable manifolds
are indispensable for precisely understanding phenomena occurring in
the micro devices.
As a way to extract light from optical cavities, the chaos-assisted
directional emission adds a new idea to the conventional theories
\cite{Siegman86}, and it would be of interest to apply this idea to
design practical devices.

\acknowledgements
Shinohara and Hentschel acknowledge financial support from the DFG
research group 760 ``Scattering Systems with Complex Dynamics'' and
the DFG Emmy Noether Program.
\end{document}